Accepted for presentation at EMBC 2020# Pain and Physical Activity Association in Critically Ill Patients*

Anis Davoudi, Tezcan Ozrazgat-Baslanti, Patrick J. Tighe, Azra Bihorac, Parisa Rashidi*Abstract—* Critical care patients experience varying levels of pain during their stay in the intensive care unit, often requiring administration of analgesics and sedation. Such medications generally exacerbate the already sedentary physical activity profiles of critical care patients, contributing to delayed recovery. Thus, it is important not only to minimize pain levels, but also to optimize analgesic strategies in order to maximize mobility and activity of ICU patients. Currently, we lack an understanding of the relation between pain and physical activity on a granular level. In this study, we examined the relationship between nurse assessed pain scores and physical activity as measured using a wearable accelerometer device. We found that average, standard deviation, and maximum physical activity counts are significantly higher before high pain reports compared to before low pain reports during both daytime and nighttime, while percentage of time spent immobile was not significantly different between the two pain report groups. Clusters detected among patients using extracted physical activity features were significant in adjusted logistic regression analysis for prediction of pain report group.

*Clinical Relevance—* This study shows that patient-reported pain scores and objectively measured physical activity are related in critical care patients.## I. INTRODUCTION

More than half of the patients in the critical care settings experience moderate to severe pain during their stay in the Intensive Care Unit (ICU) [1]. Pain in critically ill patients is often treated using opioid analgesics, which in some studies have been shown to increase the short-term and long-term risks associated with conditions such as delirium [2]. To properly manage pain, ICU nurses assess pain every hour using pain scales such as the Visual Analog Scale (VAS) for communicative patients, and the Behavioral Pain Scale (BPS), Critical Care Pain Observation (CPOT) and Non-Verbal Pain Scale (NVPS) for nonverbal patients [3-5]. While helpful for assessing pain intensity, existing scales and pain management strategies commonly focus on minimizing pain intensity rather than enhancing functional recovery [6].

Pain intensity also can impact physical activity level in critically ill patients, and which in turn can affect the pace of recovery. Currently, we lack an understanding of the relation between pain and physical activity on a granular level. At present, there is not a standard of care for collecting objective physical activity data in ICU patients beyond simple observational scales [7]. A few previous studies have examined the feasibility of using wearable accelerometer devices for collecting physical activity data with respect to indices such as delirium and sedation/agitation in ICU patients [8, 9], as well as for evaluating the feasibility of automating of pain medications administration [10]. However, the relationship between pain and physical activity has not been investigated thoroughly in this population, especially based on objective pain assessments. In this study, we investigated the relationship between physical activity and pain intensity. Physical activity is objectively measured using a wrist-worn accelerometer device and pain intensity is assessed by ICU nurses using the Defense & Veterans Pain Rating Scale (DVPRS) [11] pain scale.

## II. METHODS

### A. Patient Recruitment

This study was approved by the University of Florida (UF) Institutional Review Board (IRB 201400546). We recruited from a pool of surgical ICU patients who were expected to stay in the ICU for more than 24 hours. Recruited patients provided written informed consent to participate in the study. Recruited patients wore ActiGraph GT3X+ [12] accelerometer devices on their righthand wrist for the duration of their stay in the ICU or up to seven days, whichever preceded. Accelerometer data were downloaded from the devices and converted using the ActiLife toolbox [13]. We retrieved the patients' relevant electronic health records (EHR) information including their demographics and hospital admission data from the UF's Integrated Data Repository service. Patients' self-reported pain scores were recorded by the nurses every hour in EHR system using the DVPRS scale.

### B. Analysis

To analyze physical activity captured by the accelerometers, we extracted four statistical features from the vector magnitude of activity counts in the cartesian axes of accelerometer data. Extracted features include a) average, b) standard deviation, c) maximum, and d) percentage of time spent immobile. For feature extraction, we used 15-minute time windows, starting 30 minutes before pain assessment. We excluded the 15 minutes right before pain assessment to minimize the effect of patient movement in presence of the nurse as a result of medical procedures (Figure I). For the remainder of this analysis, daytime was defined as 7am-7pm and nighttime is defined as 7pm-7am based on routine nurse shift changes. Next, we investigated the relationship between features extracted from accelerometer data with pain scores (0-10) using linear regression analysis. We also used unsupervised clustering to detect physical activity profiles using activity features. We used Hopkins statistics to validate the clustering tendency in the dataset. Then, K-means method

*This work is supported by NSF CAREER 1750192 (PR), NIH/NIBIB 1R21EB027344 (PR, AB), NIGMS R01GM110240 (PR, AB), and R01 GM114290 (PJT, PR).A. Davoudi is with the University of Florida, Gainesville, FL 32611 USA (email: anisdavoudi@ufl.edu).

T. Ozrazgat-Baslanti is with the University of Florida, Gainesville, FL 32611 USA (email: tezcan@medicine.ufl.edu).

P. J. Tighe is with the University of Florida, Gainesville, FL 32611 USA (email: ptighe@medicine.ufl.edu).

A. Bihorac is with the University of Florida, Gainesville, FL 32611 USA (email: abihorac@medicine.ufl.edu).

P. Rashidi is with the University of Florida, Gainesville, FL 32611 USA (corresponding author, phone: ; fax: ; email: parisa.rashidi@ufl.edu).

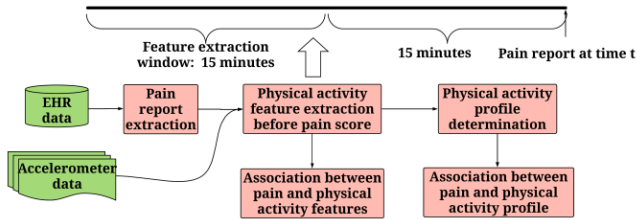

Figure 1 Analysis workflow.

was used for clustering activity features and the optimal number of clusters was determined using Silhouette metric [14-16]. We examined the effect of pain on physical activity profiles of the ICU patients based on clustering. Finally, we used logistic regression to study the effect of pain scores on physical activity profiles when adjusting for demographics (age, gender, height and weight) and the time of pain record (daytime, nighttime). Statistical significance of differences was determined using Student's t-test for continuous variables and chi-squared test for categorical variables. The analyses in this study were performed using R 3.6.2.

### III. RESULTS

To examine the association between pain and physical activity, we recruited 92 patients at University of Florida, Shands hospital, medical and surgical ICUs. We excluded patients who were later transferred from ICU and thus had incomplete data (N = 10) and patients who were unable to wear the ActiGraph device on their wrist (N = 12) and only included data from those with complete wrist data. Figure II shows the enrolled patients' cohort. For the remaining patients (N=59), 3797 pain assessments were recorded during data collection period. Out of these, 940 pain records were labeled as patients being asleep, 42 were blank pain records, and one point was outlier higher than the acceptable range of movement activity count; leaving 2814 pain records from 58 patients. Table I shows the distribution of demographic and clinical variables for the patients included in the analyses. The included and excluded patient groups were not significantly different in terms of demographics and clinical variables. Table II shows the distribution of pain scores. Pain scores were binned into two groups: (1) Mild pain: pain scores: 0-4, (2) Severe pain: pain scores: 5-10. There were 1976 (70.2%) mild pain scores and 838 (29.8%) severe pain scores. Average, standard deviations and maximum of activity counts were significantly different between the two pain classes for all pain assessments, during daytime, and during nighttime. Percentage of time spent immobile was not found to be significant. Analysis of data using Silhouette metric determined the optimal number of clusters to be two clusters based on using physical activity features with K-means algorithm (Figure III). Average, standard deviation and maximum activity counts were significantly different between mild and severe pain reports during day and night, during the day, and during the night (Table III). Multivariate logistic regression analysis for modeling pain severity using physical activity features showed only percentage of time spent immobile to be significant during daytime when adjusting for age, BMI, time of pain report, and gender (Table IV). The physical activity features were also significantly different between the two clusters; where samples in cluster 2 on average show lower average of physical activity count and lower maximum values, with significantly higher percentage spent immobile (Table V). The samples in cluster 2 are also associated with lower pain scores (p-values <0.0001 for during day and night, during daytime, and during nighttime) (Figure IV). Repeating the analysis for determining pain score classes using physical activity profiles defined using clustering instead of physical activity features shows that physical activity profiles were significant in the multivariate logistic regression analysis when adjusting for age, gender, demographics and time of day of the pain report, which remained significant for nighttime pain reports but not for daytime pain reports (Table VI); while it was significant for unadjusted analysis (p-value: <0.0001), during daytime (p-value: 0.0286), and during nighttime (p-value: <0.0001) (Figure IV).

TABLE I. PATIENT CHARACTERISTICS

| Variable | Patients in the analysis (58) | Patients not included (34) | p-value |
|---|---|---|---|
| Age, mean (sd) | 61.2 (17.6) | 61.7 (14.0) | 0.8729 |
| Gender: Female N (%) | 24 (41.4) | 8 (23.5) | 0.1315 |
| Weight, mean (sd) in kg | 84.3 (24.7) | 85.9 (31.9) | 0.8032 |
| Height, mean (sd) in cm | 170.9 (11.8) | 174.5 (10.2) | 0.1272 |
| Race, White N (%) | 53 (91.4) | 26 (76.5) | 0.09462 |
| ICU length of stay (hours) | 17.0 (15.6) | 14.0 (11.9) | 0.2996 |
| Hospital length of stay (hours) | 22.4 (18.0) | 21.3 (16.1) | 0.7703 |
| Delirium, Positive N (%) | 22 (37.9) | 18 (52.9) | 0.5229 |

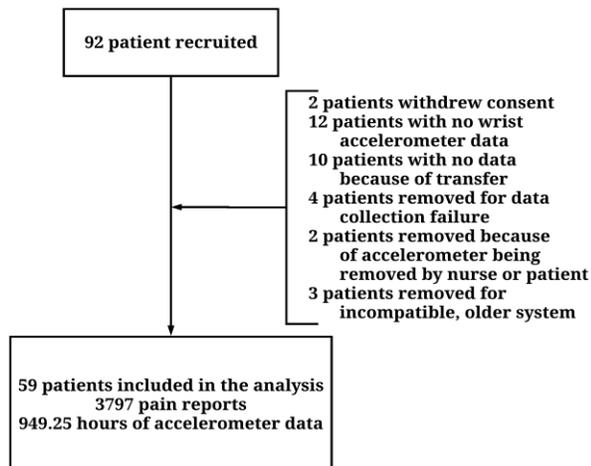

Figure 2 Data cohort selection.

TABLE II. DISTRIBUTION OF PAIN SCORES.

| Pain classes | Pain scores | Number of samples |
|---|---|---|
| Patient asleep | - | 940 |
| Mild pain: 1976 | 0 | 1247 |
| | 1 | 62 |
| | 2 | 206 |
| | 3 | 239 |
| | 4 | 222 |
| Severe pain: 838 | 5 | 194 |
| | 6 | 192 |
| | 7 | 200 |
| | 8 | 150 |
| | 9 | 68 |
| | 10 | 34 |

TABLE III.  DISTRIBUTION OF PHYSICAL ACTIVITY FEATURES BY SEVERITY OF PAIN SCORES.

| | Day and night | | | Daytime | | | Nighttime | | |
|---|---|---|---|---|---|---|---|---|---|
| Variable, mean (SD) | Low pain | High pain | p-value | Low pain | High pain | p-value | Low pain | High pain | p-value |
| Mean | 183.2 (397.2) | 256.6 (466.0) | **<0.0001** | 207.8 (429.6) | 277.5 (517.8) | **0.0097** | 141.3 (330.8) | 230.4 (390.4) | **0.0002** |
| Standard deviation | 210.8 (352.2) | 286.1 (413.1) | **<0.0001** | 231.0 (359.3) | 305.4 (444.6) | **0.0012** | 176.2 (337.3) | 261.9 (368.8) | **0.0002** |
| Maximum | 702.2 (1176.3) | 947.0 (1365.3) | **<0.0001** | 771.5 (1217.5) | 1025.0 (1483.0) | **0.0010** | 583.9 (1093.3) | 848.7 (1195.6) | **0.0004** |
| Percentage of time spent immobile | 57.3 (33.1) | 57.6 (34.4) | 0.7976 | 53.4 (32.7) | 55.5 (33.8) | 0.2351 | 63.9 (32.6) | 60.3 (35.1) | 0.0957 |

TABLE IV.  LOGISTIC REGRESSION FOR DETERMINING PAIN SCORES USING FEATURES EXTRACTED FROM ACCELEROMETER DATA ADJUSTED FOR AGE, TIME OF DAY FOR THE PAIN REPORT, GENDER, AND BMI.

| | Day and night | | Daytime | | Nighttime | |
|---|---|---|---|---|---|---|
| Variable | OR (95% CI) | p-value | OR (95% CI) | p-value | OR (95% CI) | p-value |
| Mean | 1.00 (1.00, 1.00) | 0.7109 | 1.00 (1.00, 1.00) | 0.4005 | 1.00 (1.00, 1.00) | 0.4721 |
| Max | 1.00 (1.00, 1.00) | 0.9365 | 1.00 (1.00, 1.00) | 0.5701 | 1.00 (1.00, 1.00) | 0.5094 |
| SD | 1.00 (1.00, 1.00) | 0.3027 | 1.00 (1.00, 1.00) | 0.7031 | 1.00 (1.00, 1.00) | 0.4423 |
| Percentage of time spent immobile | 0.80 (0.59, 1.09) | 0.1601 | **0.63 (0.42, 0.94)** | **0.0247** | 1.06 (0.66, 1.70) | 0.8097 |
| Day time: Nighttime | 0.74 0.63 (0.88) | **5.7 e-4** | - | - | - | - |
| Age | 1.02 (1.01, 1.02) | **2.6 e-12** | 1.02, (1.02, 1.03) | **1.5 e-9** | 1.01 (1.01, 1.02) | **0.0001** |
| BMI | 0.98 (0.97, 0.99) | **1.7 e-5** | 0.97 (0.96, 0.98) | **2.2 e-5** | 0.99 (0.97, 1.00) | 0.1045 |
| Gender: Male | 0.94 (0.80, 1.12) | 0.5002 | **0.88 (0.70, 1.10)** | 0.2582 | 1.02 (0.78, 1.33) | 0.8825 |

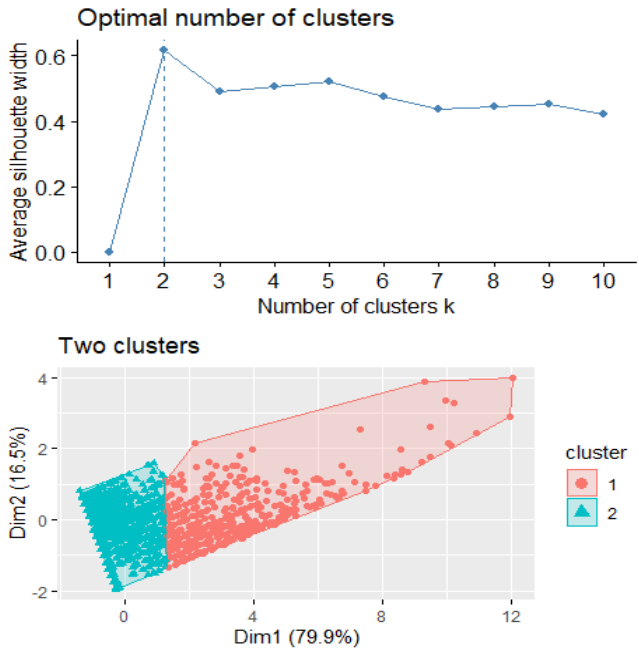

Figure 3 Optimal number of clusters using Silhouette metric and K-means clustering; data distribution in two clusters.

TABLE V.  DISTRIBUTION OF PHYSICAL ACTIVITY FEATURES AMONG DETECTED PHYSICAL ACTIIVTY PROFILES.

| Variable, mean (SD) | Cluster 1 | Cluster 2 | p-value |
|---|---|---|---|
| Mean | 913.3 (605.8) | 56.1 (94.2) | **<0.0001** |
| Standard deviation | 895.2 (415.4) | 94.0 (143.1) | **<0.0001** |
| Maximum | 2972.9 (1386.1) | 312.9 (478.4) | **<0.0001** |
| Percentage of time spent immobile | 19.9 (19.2) | 65.3 (30.3) | **<0.0001** |

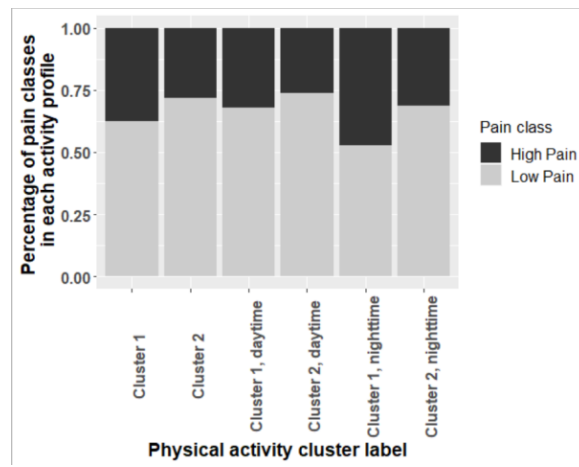

Figure 3 Percentage of pain reports' classes in each physical activity profiles for day and night, during daytime, and during nighttime.

## IV. DISCUSSION

In this study, we investigated the relationship between physical activity as objectively measured using wearable accelerometer devices and patient-reported pain in the critical care settings. Our results showed that pain and physical activity are related when adjusting for relevant demographic variables. We examined the relationship between pain and physical activity separately for pain reports during daytime and nighttime, since pain might have a negative effect on patients' physical activity profiles, and the desirable activity profiles during the night and during the day are different. During the daytime, desirable physical activity profile for the critical care patients consists of higher and more frequent activity. During the night, the desirable physical activity consists of low and limited physical activity, with long periods of continuous sleep and rest. As expected, percentage of time spent immobile during the night is higher than during the day; however, the results agree with previous observations about lack of nightly rest among ICU patients.

TABLE VI. LOGISTIC REGRESSION FOR DETERMINING PAIN SCORES USING PHYSICAL ACTIVITY PROFILES, ADJUSTED FOR AGE, TIME OF DAY FOR THE PAIN REPORT, GENDER, AND BMI.

|  | Day and night | | Daytime | | Nighttime | |
|---|---|---|---|---|---|---|
| Variable | OR (95% CI) | p-value | OR (95% CI) | p-value | OR (95% CI) | p-value |
| Physical activity profile: Cluster 2 (vs Cluster 1) | 1.49 (1.21, 1.84) | **1.8 e-4** | 1.28 (0.97, 1.68) | 0.0746 | 1.91 (1.36, 2.67) | **1.8 e-4** |
| Age, per 1 unit increase | 1.02 (1.01, 1.02) | **<0.0001** | 1.02 (1.01, 1.03) | **<0.0001** | 1.01 (1.01, 1.02) | **<0.0001** |
| Daytime: Nighttime (vs. Daytime) | 0.74 (0.63, 0.87) | **4.1 e-4** | - | - | - | - |
| Gender: Male (vs Female) | 0.96 (0.81, 1.13) | 0.6169 | 0.91 (0.96, 0.98) | 0.4265 | 1.00 (0.78, 1.30) | 0.9718 |
| BMI, per 1 unit increase | 0.98 (0.97, 0.99) | **<0.0001** | 0.97 (0.96, 0.98) | **<0.001** | 0.98 (0.97, 1.00) | 0.0591 |

One of the main limitations of the study is not including other clinical variables such as the underlying medical condition, admission diagnosis, length of ICU/hospital stay, history of physical and cognitive impairment, physical constraints in the ICU, and administration of any sedative or muscle relaxer medications. Another limitation of the work is accurate detection of weartime for the wearable accelerometers. Current weartime detection algorithms depend on movement and are not suitable and validated for critical care settings. Validation of the reported results in more diverse and larger datasets will contribute to the efforts directed at automating detection of pain in the ICU patients, as well as better understanding the effect of pain and pain relief sedatives on physical activity in the critical care settings. Larger datasets may also allow for more stratified analysis of the relationship between pain and physical activity.